\pdfoutput=1
\documentclass[10pt]{article}

\usepackage{amsmath}
\usepackage{amssymb}

\usepackage{graphicx}

\usepackage{cite}

\usepackage{color}

\usepackage[labelfont=bf,labelsep=period,justification=raggedright]{caption}

\date{}

\usepackage[pagewise]{lineno}
\makeatletter
\renewcommand{\@biblabel}[1]{\quad#1.}
\makeatother

\begin{document}

\begin{flushleft}
{\Large
\textbf{Initial conformation of kinesin\textquoteright s neck linker}
}
\\
Yi-Zhao Geng$^{1}$,
Qing Ji$^{2,3,\ast}$,
Shu-Xia Liu$^{2}$,
Shiwei Yan$^{1,4,\ast}$
\\
\bf{1} College of Nuclear Science and Technology, Beijing Normal University, Beijing 100875, China
\\
\bf{2} Institute of Biophysics, Hebei University of Technology, Tianjin 300401, China
\\
\bf{3} School of Science, Hebei University of Technology, Tianjin 300401, China
\\
\bf{4} Beijing Radiation Center, Beijing 100875, China
\\
$\ast$ E-mail: jiqingch@hebut.edu.cn (QJ); yansw@bnu.edu.cn (SY)
\end{flushleft}

\section*{Abstract}
How ATP binding initiates the docking process of kinesin\textquoteright s neck linker is a key question in understanding kinesin mechanism. It is believed that the formation of an extra turn structure by the first three amino acids of neck linker (LYS325, THR326, ILE327 in 2KIN) is crucial for initiating the docking process. But the initial conformation of neck linker (specially the three amino acids of the extra turn) and the neck linker docking initiation mechanism remain unclear. By using molecular dynamics method, we investigate the initial conformation of kinesin's neck linker in the docking process. We find that, in the initial state of NL docking process, NL still has interactions with $\beta$0 and forms a conformation similar to the \textquoteleft\textquoteleft cover-neck bundle\textquoteright\textquoteright~structure proposed by Hwang et al.~[\emph{Structure} 2008, 16(1): 62-71]. From this initial structure, the docking of the \textquoteleft\textquoteleft cover-neck bundle\textquoteright\textquoteright~structure can be achieved. The motor head provides a forward force on the initial cover-neck bundle structure through ATP-induced rotation. This force, together with the hydrophobic interaction of ILE327 with the hydrophobic pocket on the motor head, drives the formation of the extra turn and initiates the neck linker docking process.

\section*{Introduction}
Conventional kinesin (kinesin-1) \cite{2004-Lawrence-JCB} is an ATP-powered microtubule-based dimeric motor protein responsible for mitosis and organelle transportation in cells \cite{2000-Vale-Science,2003-Vale-Cell,1993-Svoboda-Nature,2009-Hirokawa-NatureRev.Mol.CellBiol}. Kinesin\textquoteright s two identical motor heads are connected by a coiled-coil stalk and two neck linkers (NLs). The two heads switch their positions on microtubule in a hand-over-hand manner \cite{2000-Vale-Science,2004-Yildiz-Science,2003-Kaseda-Nat.CellBiol,2003-Asbury-Science}. In each walking cycle, the two NLs have large conformational changes and the docking process of the leading head\textquoteright s NL to the motor domain constitutes a crucial part of kinesin\textquoteright s force generation (the power stroke) \cite{1999-Rice-Nature,2000-Case-Curr.Biol.,2002-Sindelar-Nat.Struct.Biol,2006-Tomishige-NSMB}. The power stroke by NL docking is triggered by ATP binding, rather than ATP hydrolysis, in the presence of microtubule \cite{1999-Rice-Nature,2002-Sindelar-Nat.Struct.Biol,2008-Kikkawa-TrendsinCellBiology}. The force generated during the NL docking process arises from the contributions of various interactions among amino acids (AAs) of kinesin and other molecules. The NL docking force-generation mechanism should be understood on structural basis with correct initial conformation of NL.

Kinesin\textquoteright s NL is a short peptide ($\sim$ 14 AAs, LYS325 - THR338 in 2KIN) which consists of three parts, including the first three AAs of extra turn, $\beta$9 and $\beta$10 between $\alpha$6 of the motor domain and $\alpha$7 of the coiled coil \cite{2000-Vale-Science,1999-Rice-Nature,2002-Sindelar-Nat.Struct.Biol,1997-Sack-Biochemistry}. It has been revealed that NL of the leading head has a rearward positioning as its initial conformation in its docking process and NL of the trailing head has a forward extension along the length of the motor domain as depicted in, for example, Fig.~2a of Ref.~\cite{2006-Tomishige-NSMB}. The docking process of NL is crucial for force generation and replacement of NL with a designed random coil results in the decrease of microtubule velocity \cite{2000-Case-Curr.Biol.}. NL docking is accomplished in three steps: (1) the first three AAs forming an extra turn structure, (2) $\beta$9 docking to the motor domain and then (3) $\beta$10 docking to the motor domain. Simulations and experiments by Hwang et al.~\cite{2008-Hwang-Structure}and Khalil et al.~\cite{2008-Khalil-PNAS} have shown that the second step of NL docking is achieved by forming a cover-neck bundle (CNB), a special $\beta$-sheet formed by the N-terminal $\beta$0 and $\beta$9. The MD simulations by Hwang et al.~\cite{2008-Hwang-Structure} start from a conformation within which the first three AAs of NL already form an extra turn. Then there remain several questions, such as: what is the initial conformation of $\beta$0 and NL before formation of extra turn? what is the mechanical pathway for the formation of the extra turn?

To address these questions, we perform a series of molecular dynamics (MD) simulations. We find that, in NL\textquoteright s initial undocked state, NL and $\beta$0 are not completely \textquoteleft\textquoteleft free\textquoteright\textquoteright. Instead, there is a group of hydrogen bonds between them, i.e., CNB structure already exists at the beginning of NL\textquoteright s docking process. The docking of CNB proposed by Hwang et al.~\cite{2008-Hwang-Structure} can be achieved from this initial CNB structure. Based on the ATP- and ADP-state crystal structures of kinesin, we also find that the motor head rotation induced by ATP binding provides a forward force on the initial CNB structure. This force, together with the hydrophobic interaction between ILE327 and the hydrophobic pocket, drives the formation of the extra turn.

\section*{Methods}
The structure we use for modeling is 2KIN \cite{1997-Sack-Biochemistry} and 1BG2 \cite{1996-Kull-Nature}. 2KIN is the X-ray structure of rat kinesin. It has docked NL and is suggested to be in ATP-like state. 1BG2 is human kinesin of ADP state and lacks NL. To save computational resource, we deleted 2KIN\textquoteright s $\alpha$7 (the residues after ALA339), ADP and SO4. The motor domain is surrounded by explicit water and the spherical boundary condition is used. 2KIN does not have coordinates for the main part of L11 loop. We connecte the two ends of L11 (VAL239 and ASN256 in 2KIN) directly. Because our model does not include microtubule, we fix the C$_\alpha$ atoms of GLU158 at $\beta$5, SER260 and GLU271 at $\alpha$4 (AA numbering of 2KIN) to mimic the microtubule-bound state. The MD simulations are performed by using NAMD (version 2.9) \cite{2005-Phillips-J.Comput.Chem} with force field CHARMM \cite{1998-MacKerell-JPC} at the temperature 310K. The integration time step is 2 fs. The software we use for modeling and molecular graphics is VMD (version 1.9.1) \cite{1996-Humphrey-JMG}.  Na$^{+}$ and Cl$^{-}$ are added to ensure an ionic concentration of 150 mM and zero net charge \cite{2011-Li-Biochemistry}. The non-bonded Coulomb and vdW interactions are calculated with a cutoff using a switching function starting at a distance of 13 {\AA} and reaching zero at 15 {\AA}. The water model we use is TIP3P \cite{1983-Jorgensen-JCP}. In the simulations, the protein-water system is minimized for 30000 steps MD simulation to remove bad contacts.

\section*{Results and Discussion}
We construct three different conformations for NL and $\beta$0 of the leading head in the ATP-waiting state and then run MD simulations to check which one reasonably reflect the real situation. The construction is based on two crystal structures of kinesin, 2KIN \cite{1997-Sack-Biochemistry} and 1BG2 \cite{1996-Kull-Nature}. 2KIN is of the ATP-bound state conformation and 1BG2 is of the ADP-bound state conformation \cite{2007-Sindelar-JCB,2010-Sindelar-PNAS}. Because the nucleotide-free state conformation is similar with that of the ADP-bound state \cite{2007-Sindelar-JCB,2010-Sindelar-PNAS}, we use the crystal structure of 1BG2 as the nucleotide-free state conformation in this investigation and call both ADP-state and nucleotide-free state conformations as ADP-state conformation.

\begin{figure}[htbp]
\begin{center}
\includegraphics[width=4in]{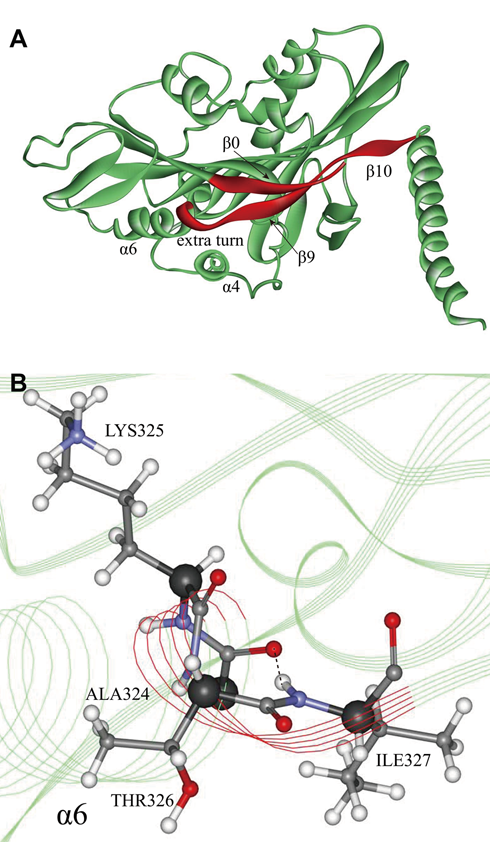}
\end{center}
\caption{
{\bf The CNB and extra turn in the ATP state conformation of kinesin.} (A) The motor head structure in the ATP state conformation. NL and $\beta$0, which form a CNB structure, are depicted in red. (B) The closed extra turn structure in the ATP state conformation. The alpha carbons are depicted by bigger spheres and the hydrogen bond by dotted line. The side chain of ALA324 is omitted with only the alpha carbon and the peptide plane remained. Molecular drawings are produced by using Discovery studio 3.5 visualizer.
}
\label{1}
\end{figure}

The docking of NL starts from the formation of extra turn. When NL is docked, its first three AAs, LYS325, THR326 and ILE327 (in 2KIN), form an extra turn structure (Fig.~\ref{1}). LYS325 and ILE327 are two of the most conservative AAs of kinesin. THR326 can be replaced by a serine, both of which have a hydroxide group \cite{2009-Hariharan-CellMol.Bioengineering}. The signature of the extra turn formation is the formation of a hydrogen bond between the backbone hydrogen of ILE327 and the backbone oxygen of ALA324, as depicted in Fig.~\ref{1}B. The condition for the formation of this hydrogen bond is that ILE327 is docked in the hydrophobic pocket formed by six hydrophobic AAs (including ILE9, ILE266, LEU269,ALA270, LEU292 and ALA324 in 2KIN) of the motor domain. When ILE327 is outside of the hydrophobic pocket, the extra turn is open.

\subsection*{CNB structure cannot be achieved from a conformation with free $\beta$0}
\begin{figure}[htbp]
\begin{center}
\includegraphics[width=4in]{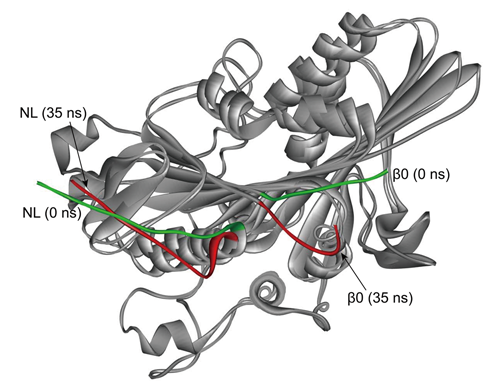}
\end{center}
\caption{
{\bf Overlaid snapshots of initial and finial state conformations of 35 ns MD simulation with free $\beta$0.} NL and $\beta$0 of the initial state conformation are shown in green color and that of the finial state are shown in red.
}
\label{2}
\end{figure}

In the ATP waiting state of kinesin, the leading head is in nucleotide-free state and the trailing head is located 8 nm behind the leading head \cite{2009-Toprak-PNAS,2009-Asenjo-PNAS,2008-Yildiz-Cell}. The leading head\textquoteright s NL points to the microtubule\textquoteright s minus end in the NL initial undocked state. However, the conformation of $\beta$0 in the initial state is still unclear. In some references, $\beta$0 is assumed to be free \cite{2006-Tomishige-NSMB,2007-10-Hyeon-PNAS,2008-Hwang-Structure,2008-Khalil-PNAS}. To check whether the CNB structure in the second step of NL docking could be achieved from this initial conformation, we construct a initial structure for NL docking from the crystal structure of 2KIN. In this structure, NL has a backward orientation and $\beta$0 is free. Extra turn of this structure is in the open state (Fig.~\ref{2}). Taking this conformation as the initial state of NL docking, we perform an MD simulation.

The result of a 35 ns simulation shows that NL has no tendency of docking to the motor domain and $\beta$0 has no tendency of forming CNB structure with NL either (Fig.~\ref{2}). This result indicates that CNB structure cannot be achieved from a conformation with free $\beta$0, even though $\alpha$4 and $\alpha$6 keep their relative positions in the ATP-state conformation and there is no obstruction of $\alpha$4 to NL docking.

Form Fig.~\ref{2}, it is seen that $\beta$0 should have a large backward conformational change from its free state to form CNB structure with NL. But, there is no force to drive $\beta$0 to achieve such a large conformational change. Therefore, the structure with free $\beta$0 is not the correct initial structure of NL docking.
\begin{figure}[htbp]
\begin{center}
\includegraphics[width=4in]{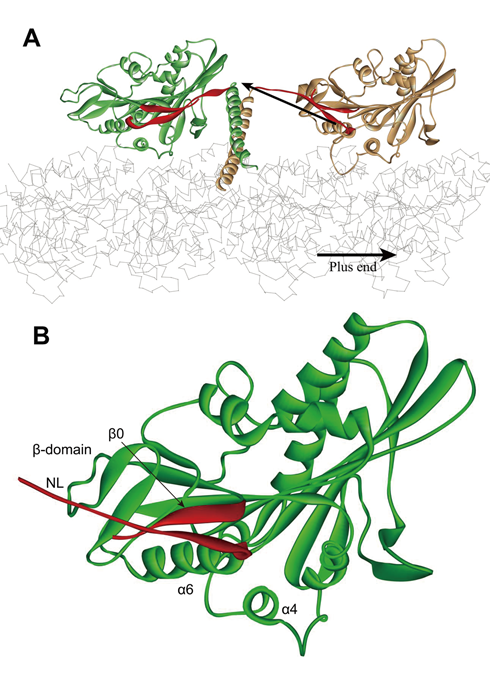}
\end{center}
\caption{
{\bf Backward CNB structure based on ATP state structure of 2KIN.} (A) Two head bound conformation of kinesin on microtubule. The direction of the unbinding force is indicated by the arrow. The magnitude of the force is 160 pN in the NL unbinding simulation. (B) The resultant backward conformation of NL and $\beta$0, which is also a CNB structure. $\beta$0 locates between NL and $\beta$-domain.
}
\label{3}
\end{figure}

\subsection*{Without extra turn formation, a backward CNB structure cannot properly dock to the motor domain}
To obtain the initial conformation of NL and $\beta$0 of the leading head, we unbind the docked NL of 2KIN by exerting a backward force at the C-terminus of NL in MD simulation. The direction of this force is depicted in Fig.~\ref{3}A. The direction of the unbinding force is set based on the following facts. In the hand-over-hand movement of kinesin, the two motor heads walk along the same protofilament on the microtubule surface \cite{2003-Skiniotis-EMBOJ.}. The force acting on the leading head\textquoteright s NL arises from the inter-head tension between the two motor heads \cite{2008-Yildiz-Cell,2011-Clancy-Nat.Struct.Mol.Biol,2002-Rosenfeld-JBC,2003-Rosenfeld-JBC,2003-Uemura-NSB}. Therefore, this force should point to the minus end of microtubule as depicted in Fig.~\ref{3}A. The magnitude of the inter-head tension is estimated as 15 - 35 pN \cite{2009-Hariharan-CellMol.Bioengineering}. Due to the limitation of computational capacity, we use 160 pN backward force to unbind the NL. In our simulation, this force is the smallest force which can unbind NL within tens of nanoseconds. This force is much weaker than that used in the simulation of Hwang et al.~($>$ 400 pN). After 13 ns simulation, NL arrives at a backward pointing position similar to the NL position of the motor head in Fig.~\ref{2}. In the mean time, $\beta$0 still keeps a CNB structure with NL. In the whole unbinding process, four hydrogen bonds keep between $\beta$0 and NL (two hydrogen bonds between VAL331 and ALA5 and another two between CYS7 and ASN329). This is a very interesting result beyond expectation because $\beta$0 and NL are exposed to the surrounding water during the whole pulling process. In this backward-pointing CNB structure, $\beta$0 locates between NL and the $\beta$-domain ($\beta$2a, $\beta$2b and $\beta$2c) and the extra turn is open (Fig.~\ref{3}B).
\begin{figure}[htbp]
\begin{center}
\includegraphics[width=4in]{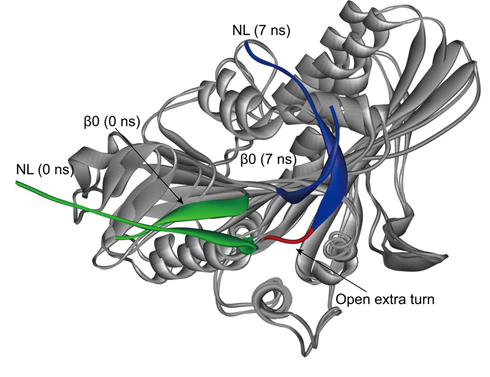}
\end{center}
\caption{
{\bf Overlaid snapshots of initial and finial state conformations of kinesin in the MD simulation with backward-pointing CNB based on ATP state structure.} NL and $\beta$0 of the initial state are shown in green and that of the finial state are shown in blue. The open extra turn of the finial state is shown in red.
}
\label{4}
\end{figure}

A series of MD simulations is run with this structure as the initial state, we never, however, get a conformation with rightly docked NL. In these simulations, CNB structure has obvious docking tendency which shows that CNB structure has the mechanical property of bending forward \cite{2008-Hwang-Structure}. But, in these simulation results, the first three AAs of NL never form a correct extra turn structure, i.e., the extra turn keeps open (see Fig.~\ref{4}). Without extra turn formation, the residues of $\beta$9 cannot be in register with their docking sites in the motor domain. Therefore, without extra turn formation, a backward CNB structure cannot dock to the motor domain properly.

In the above simulations, the initial conformation of motor head is obtained based on the crystal structure of 2KIN, which is in ATP-like state. From this conformation, extra turn cannot form, as is explained in the next section. Therefore, the backward-pointing CNB based on ATP state structure cannot dock to the motor domain in a right way.

\subsection*{Initial conformation of NL and $\beta$0 should be a rearward pointing CNB based on nucleotide-free state structure}
\begin{figure}[htbp]
\begin{center}
\includegraphics[width=4in]{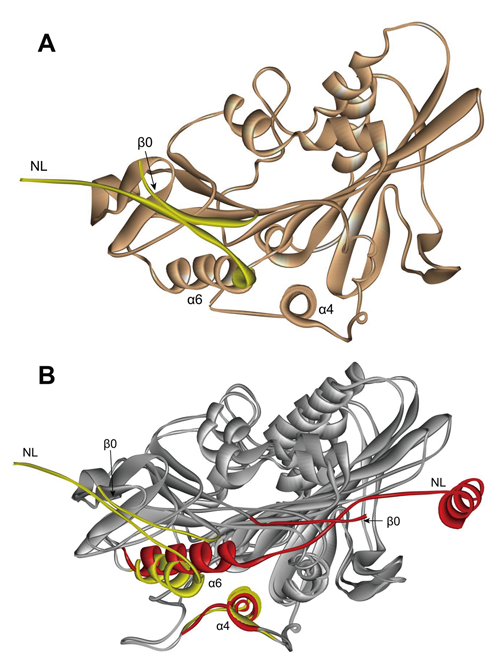}
\end{center}
\caption{
{\bf The initial backward-pointing CNB based on ADP state structure of kinesin.} (A) Kinesin\textquoteright s ADP state initial conformation for NL docking with NL and $\beta$0 constituting a backward-pointing CNB structure. (B) Superposition of kinesin\textquoteright s ATP state structure (2KIN) and ADP state initial structure (based on 1BG2). $\alpha$4s of the two structures are kept coincided. $\alpha$6 of the ATP state structure (shown in red) has an inward rotation relative to its initial conformation (shown in yellow).
}
\label{5}
\end{figure}

As is well known, NL docking is induced by ATP binding \cite{1999-Rice-Nature}. Therefore, the correct initial conformation of NL and $\beta$0 in NL docking process should belong to the nucleotide-free state conformation rather than ATP-state conformation of kinesin. This might be the reason that the extra turn cannot be formed rightly in the simulations of above section. We thus graft the CNB structure of the above-obtained conformation from 2KIN (Fig.~\ref{3}B) onto the ADP-state structure of 1BG2, which has no NL. The grafted initial CNB structure is optimized while the C-terminal end of NL is fixed to keep NL pointing backward. The resultant conformation is shown in Fig.~\ref{5}A. As seen, this conformation also has an opened extra turn and $\beta$0 locates between the $\beta$-domain and NL. We propose that this conformation should be the right initial structure of NL and $\beta$0 in the ATP-waiting state. In this structure of kinesin\textquoteright s leading head, NL and $\beta$0 still maintain a CNB structure.

As clearly pointed out by Vale and Milligan \cite{2000-Vale-Science}, one of the most important conformational differences between ATP- and ADP-state conformations of kinesin is the relative position of $\alpha$6 with respect to $\alpha$4. In the ADP state, $\alpha$6 and $\alpha$4 locate in the same plane and the C-terminal end of $\alpha$6 is close to the C-terminal end of $\alpha$4, whereas in the ATP state, $\alpha$6 locates above $\alpha$4. This conformational difference can be seen in Fig.~\ref{5}B, where the related secondary structures of both crystal structures are superposed and $\alpha$4s of these two crystal structures are kept coincided. As seen, with the two fixed $\alpha$4s, the ATP-state conformation of $\alpha$6 has an inward rotation relative to the ADP-state conformation \cite{2000-Vale-Science,2004-Skiniotis-EMBOJ.,2006-Kikkawa-EMBOJ,2010-Parke-JBC}. This large conformational change of $\alpha$6 between the two states is caused by ATP binding induced motor head rotation. How can the rotation movement of the motor domain cause the inward rotation of $\alpha$6? From structural analysis, we find that this can be accomplished through the CNB structure. The large conformational change of $\alpha$6 is directly induced by the displacement of the N-terminal end of NL, which is connected with the motor domain through forming the CNB structure with $\beta$0. $\beta$0 is connected with $\beta$1 of the central $\beta$-sheet through the peptide bond between LYS10 ($\beta$1) and ILE9 ($\beta$0). Therefore, the rotation movement of motor domain transmit through LYS10 to the CNB structure and, then, induce the rotation of $\alpha$6.
\begin{figure}[htbp]
\begin{center}
\includegraphics[width=4in]{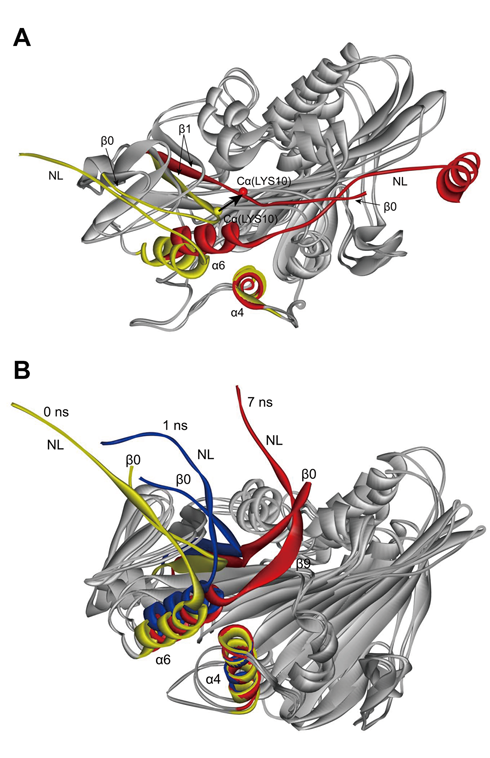}
\end{center}
\caption{
{\bf NL docking simulation starting from the ADP state initial structure of kinesin with a force exerting on the initial CNB via LYS10 of $\beta$1.} (A) The force (black arrow) exerted on C$_\alpha$ (yellow sphere) of LYS10 of $\beta$1 of the ADP-state initial structure. The direction of the force is determined by the line connecting C$_\alpha$s of LYS10s of the ADP- and ATP-state structures. The force is produced by the rotation of the motor domain which can only be transmitted to CNB via the peptide bond between ILE9 of $\beta$0 and LYS10 of $\beta$1. Therefore we exert the force on C$_\alpha$ of LYS10 in the NL docking simulation. (B) Overlaid snapshots of the simulation trajectory. The CNB at 7 ns has a close extra turn and $\beta$9 of NL has docked to the motor domain rightly.
}
\label{6}
\end{figure}

To simulate the action of the motor domain on CNB, we exert a forward force on the C$_\alpha$ of LYS10 of the grafted structure. The direction of this force is along the connection line of the two C$_\alpha$s of LYS10s in the ADP- and ATP-state conformations (the black arrow in Fig.~\ref{6}A). In our simulation, the magnitude of this forward force is 200 pN. We should mention that the magnitude of the force transmitted from the central $\beta$-sheet to CNB maybe weaker than 200 pN in the wild type kinesin. When ATP binds to motor head, it induces a group of intramolecular and intermolecular interactions which result in motor head rotation. In our simulations, the effects of these interactions are represented by a single forward force on C$_\alpha$ of LYS10. In the simulations of tens nanosecond, this force does induce the right conformational change of CNB and $\alpha$6. After 10 ns simulation, the right conformational change is obtained, i.e., the extra turn rightly forms, $\alpha$6 rotates to the position above $\alpha$4 and CNB docks to the motor head (snapshots of the simulation trajectory are shown in Fig.~\ref{6}B). This result clearly shows that the initial conformation of NL and $\beta$0 should be a rearward pointing CNB based on nucleotide-free state structure.

Above simulation results reveal that the mechanical pathway from motor head rotation to NL docking must be accomplished via the initial CNB structure. Rotation of the central $\beta$-sheet provides a forward force on the initial CNB structure through LYS10. Under the action of this force, $\beta$0 moves forward. Because $\beta$0 and NL form a CNB structure, the first three AAs of NL then has a conformational change accordingly to form the extra turn structure. In this process, $\beta$0 plays a role of force transfer element from motor head to NL by forming CNB with NL. Through analysis of the SMD trajectory (data not shown here), we find that the hydrophobic interaction between ILE327 and hydrophobic pocket of motor domain plays an important role in the formation of extra turn. The three AAs locate at the C-terminal end of $\alpha$6 and, thus, the formation of extra turn structure induces C-terminal end of $\alpha$6 rotate to the position above $\alpha$4.

In the pioneering work of Hwang et al. \cite{2008-Hwang-Structure,2008-Khalil-PNAS}, they proposed a mechanism in which the CNB does not form until after ATP binding, whereas we propose here that the CNB
is essentially stable throughout the power stroke, which is a major revision to the CNB hypothesis by Hwang et al. As shown in the above sections, CNB structure cannot be achieved from a conformation with free $\beta$0. $\beta$0 is a crucial force transfer element. Without forming an initial CNB structure, the force produced by ATP-binding induced motor head rotation cannot be transferred to NL.

\section*{Conclusions}
Initial conformation of NL is crucial for understanding the mechanical pathway from ATP-binding induced motor head rotation to NL docking. We obtain an initial structure of NL and $\beta$0, which is a backward-pointing CNB structure. Through analysis of the structural and mechanical connections between central $\beta$-sheet and CNB, we use a force acting on the root of $\beta$0 to simulate the force produced by the rotation of motor head. From the initial structure with backward-pointing CNB, we simulate the NL docking process under the action of the force. In the simulation result, the extra turn forms, $\alpha$6 rotates to the position above $\alpha$4 and CNB docks to the motor head rightly. This result proves that the mechanical pathway from rotation of central $\beta$-sheet to NL docking must be accomplished via the initial CNB structure and, thus, provides an insight into the mechanism underlying the first step of NL docking.


\section*{Acknowledgments}
One of the authors (Q.J.) thanks Hongyu Zhang for helpful discussions. The calculation was performed on the computer cluster at School of Science, Hebei University of Technology.

\bibliography{arXiv-GYZ}


\end{document}